\documentclass[10pt,letterpaper,twocolumn]{article} 
\usepackage{ol2}
\usepackage[draft]{hyperref}
\usepackage{amsmath}

\begin{document}

\twocolumn[
\title{A tunable narrowband entangled photon pair source for resonant single-photon single-atom interaction}

\author{Albrecht Haase, Nicolas Piro, J\"urgen Eschner$^*$, and Morgan W. Mitchell}

\address{ICFO - Institut de Ciencies Fotoniques,\\
Mediterranean Technology Park, 08860 Castelldefels (Barcelona), Spain\\
$^*$Corresponding author: juergen.eschner@icfo.es}

\begin{abstract}
{We present a tunable, frequency-stabilized, narrow-bandwidth source
of frequency-degenerate, entangled photon pairs. The source is based
on spontaneous parametric downconversion (SPDC) in
periodically-poled KTiOPO$_{4}$ (PPKTP). Its wavelength can be
stabilized to 850 or 854~nm, thus allowing to address two D-P
transitions in $^{40}$Ca$^+$ ions. Its output bandwidth of 22~MHz
coincides with the absorption bandwidth of the calcium ions. Its
spectral power density is 1.0 generated pairs/(s~MHz~mW).\\}
\end{abstract}

]




Entangled photon pairs have become an important resource in experiments on fundamental
quantum mechanics \cite{OuMandel}, as well as in quantum communication
\cite{Bouwmeester}, quantum computing \cite{KLM}, and quantum networks \cite{Cirac}. The
best controlled and most widely applied method to create photonic entanglement is
spontaneous parametric downconversion (SPDC) in nonlinear crystals\cite{Kwiat}. Due to
the weak phase matching conditions in small size crystals, these sources usually produce
rather broad output spectra with widths on the order of THz. For many applications this
is convenient, since the photons interact only with detectors which are not energy
selective on this scale. In recent years new SPDC sources have been developed also for
narrow band applications \cite{Wong, Zeilinger, Shapiro}, which are mostly aiming at
coupling photonic and atomic systems \cite{Kimble, Kraus}. Still, the reported bandwidths
are rather broad ($\sim100$ GHz) \cite{Wong, Zeilinger} compared to atomic transitions,
or the sources emit into multiple frequency modes \cite{Shapiro}.

In this letter we report on a photon pair source designed to permit resonant interaction
with single trapped ions \cite{Ours}. The photon frequency and bandwidth are made to
match the linewidth of the ${\rm D_{3/2}} - {\rm P}_{3/2}$ and ${\rm D_{5/2}} - {\rm
P}_{3/2}$ optical transitions in $^{40}$Ca$^+$, which are centered at 849.8 and
$854.2~$nm and have a width of $\sim 20$~MHz. As prerequisite for efficient coupling, we
achieve tunability over this range, a bandwidth reduction of six orders of magnitude
compared to standard SPDC sources, and suitable frequency stabilization. This novel
source will allow us to perform experiments on the coupling between single quantum
systems of light and matter, with applications in quantum networks.

We obtain these characteristics by optimizing the SPDC efficiency via quasi-phase
matching \cite{Boyd}, by reducing the photon pair bandwidth in a Fabry-Perot cavity
filter line, and by stabilizing the absolute photon pair frequency via active feedback on
all critical elements in the set-up.

\begin{figure*}[htb]
\centerline{\includegraphics[width=16.6cm]{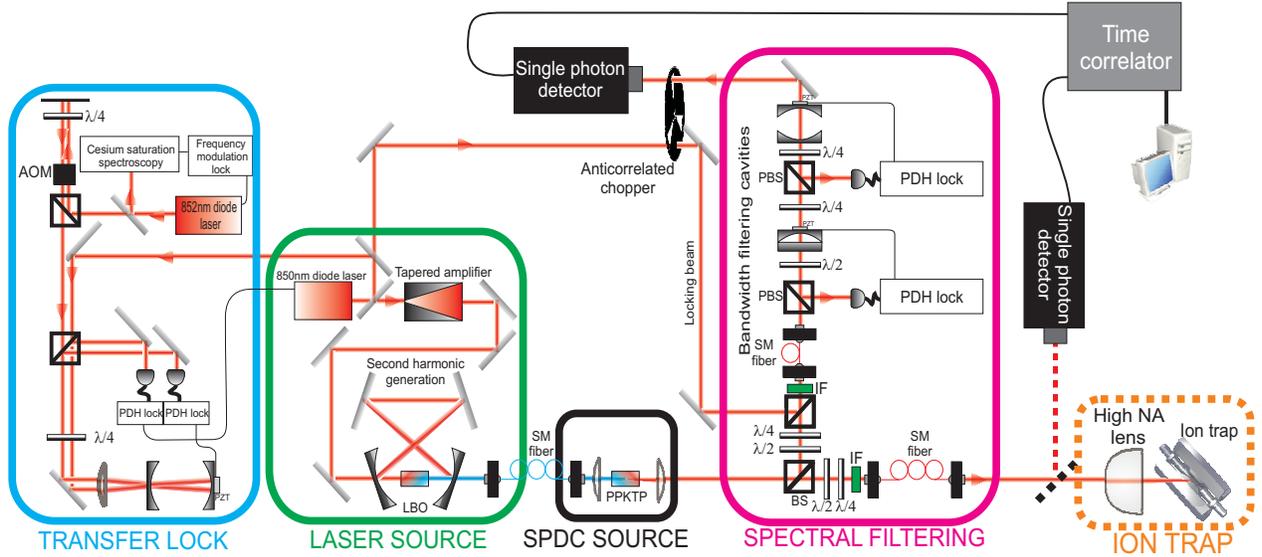}} 
\caption{Scheme of the experimental setup: The master laser is stabilized to an atomic
line (transfer lock), amplified and frequency doubled (laser source). The doubled light
is down-converted in a type II phase-matched PPKTP crystal (SPDC source). The photon
pairs are split and fiber-coupled; one of the photons is spectrally filtered (spectral
filtering). The time correlator detects the coincident pairs. Later, the unfiltered
photon will be sent to a single ion experiment. Symbols are explained in the text.}
\label{fig:setup}
\end{figure*}

The experimental setup is sketched in Figure \ref{fig:setup}. The master laser, an
extended-cavity diode laser (Toptica DL-100), provides about 30~mW of light, tunable
between 850 and 854~nm. About 90$~\%$ of it is amplified in a tapered amplifier (Toptica
TA-100) to 600~mW output power. The remaining light from the diode is used to stabilize
its frequency to a Fabry-Perot cavity of finesse 1000 and 1~MHz linewidth, using the
Pound-Drever-Hall (PDH) technique. The length of this cavity is locked to a laser at
852~nm, which itself is stabilized to the D2 line of cesium by saturation spectroscopy.
An acousto-optic modulator (AOM) in the 852~nm laser beam provides fine frequency tuning.
This transfer lock technique yields about 125~kHz absolute frequency stability of the
master laser \cite{Felix}.

The amplified master beam is frequency-doubled in a lithium tri-borate (LBO) crystal
within a bow-tie cavity (Toptica SHG-110), resonant with the master laser wavelength for
pump power enhancement, which produces around 100~mW of 425-427~nm second-harmonic light.
The resulting blue beam is sent through a single mode (SM) fiber for mode cleaning and
focussed into a second nonlinear crystal where photon pairs are created via SPDC. This
crystal is a flux-grown periodically poled KTiOPO$_4$ (PPKTP) crystal of dimensions
$20\times6\times1$ mm \cite{Carlota}. It has two independent parallel poling gratings
imprinted, with periods of 14.03 and 14.63$~\mu$m, for collinear type-II
quasi-phase-matching at a temperature of 25$^\circ$C for 849.8  and 854.2~nm,
respectively \cite{Kato}.

To optimize the focus of the pump mode, we utilize the reverse process of second-harmonic
generation (SHG), which is legitimate due to the narrow-bandwidth of the source
\cite{Morgan}. The optimum focussing parameter, the ratio of crystal length and Rayleigh
range $\xi=L/z_R=5.68$ \cite{Boyd}, requires a pump beam waist of $w_0=16.1~\mu$m. Fine
tuning and stability of the central wavelength of the photon pairs is achieved by
temperature control. The temperature of the crystal is actively stabilized with better
than 10~mK precision, and a central wavelength variation of 0.034~nm/K was measured.

The generated photon pairs are collimated and split by a non-polarizing or polarizing
beam splitter (BS or PBS), depending on whether their polarization entanglement shall be
exploited or not. In order to perform the tomographic reconstruction of the photon pair
polarization state, a set of quarter and half-waveplates and a PBS are placed in each
arm. Finally, after passing optical bandpass filters (IF) (Semrock FF01-440/40-25), the
output modes of the beam splitter are coupled into polarization maintaining single-mode
fibers. The single-photon coupling efficiencies are on the order of 42$\%$. The spectral
width of the fiber coupled photon pairs was measured to be $143 \pm 4~$GHz.

After passing through the fiber, the transmission mode of the beam splitter is coupled
into a filtering line for bandwidth reduction of the photon pairs. It consists of two
Fabry-Perot cavities placed in cascade, designed to cover the whole crystal output
spectrum and to provide a single narrow transmission window of the desired bandwidth.
Each cavity consists of two high-reflectivity mirrors (Layertec) with a measured finesse
of around 620. The first cavity of 77.5~${\mu}$m length has 3.7~GHz transmission
bandwidth, the second cavity of 10~mm length sets the final filter bandwidth of $\sim
25~$MHz. Each cavity has a measured transmission, on resonance, of 88$\%$. Together with
42$\%$ fiber coupling efficiency and 45$\%$ detector efficiency, the overall photon
detection probability in the filtered arm amounts to $\approx15\%$. .

Both cavities are individually stabilized to the master laser wavelength (850 or 854 nm),
and therefore to the desired D-P transition in $^{40}$Ca$^+$, by means of an auxiliary
beam using again the PDH technique. Conservation of energy in the downconversion process
together with the narrow frequency bandwidth of the master laser ensures that, if a
photon is transmitted through the filtering line, its partner photon will be on resonance
with the atomic transition.

Photons are detected by fiber-coupled avalanche photodiodes (Perkin Elmer SPCM-AQR-15)
with $\sim45\%$ quantum efficiency, $<$50 dark counts/s, and $<$1~ns time resolution. The
TTL detection pulses are sent to a correlation electronics module (Picoquant Picoharp
300) which is used primarily as a time discriminator to measure the delay between the
pulses in two channels. The data are analyzed with an integrated control software based
on LabviewV8.5, which furthermore controls all input and output parameters of the
experiment. This allows to remotely run complete measurement series like a full state
tomography protocol.



\begin{figure}[htb]
\centerline{\includegraphics[width=8.3cm]{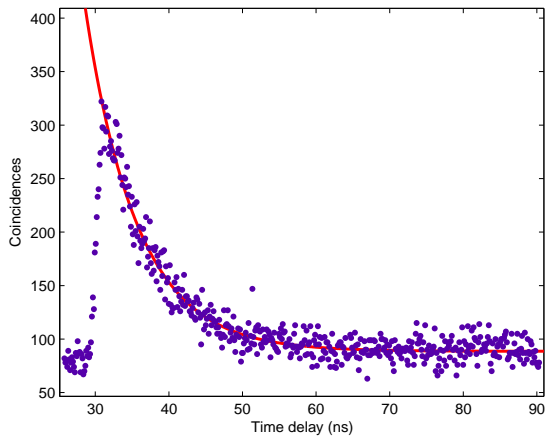}}
\caption{Time correlation between photons in the filtered and the
unfiltered arm (points), and exponential fit (line). The time origin
is shifted by an electronic delay. } \label{fig:correlations}
\end{figure}

For characterizing the spectral properties of the narrowband pair
photon source, a temporal correlation measurement was performed. The
photon pairs were split on a polarization beam splitter and detected
after filtering in one of the arms. The distribution of the time
delays between photons in the two arms is shown in
Fig.~\ref{fig:correlations}. A peak is resolved which originates
from the correlated photon pairs. It shows a characteristic
exponential decay due to the filter cavity ring-down. From the decay
time we deduce the photon pair bandwidth to be $22.4\pm0.5~$MHz, in
agreement with the measured filter cavity linewidth. The background
is caused by accidental coincidences of photons with lost partners.

The brightness of the source is obtained from the number of counts in the peak, measured
while varying the pump power. The measured rate is 4.8 pairs/(s mW). For the maximum pump
power of 70 mW this results in an extrapolated detection rate of 340 pairs/s
\cite{footnote_rate}. Taking into account the bandwidth and the detector efficiencies, we
find the spectral brightness of generated narrowband pairs to be 1.0/(s MHz mW).

\begin{figure}[htb]
\centerline{\includegraphics[width=8.3cm]{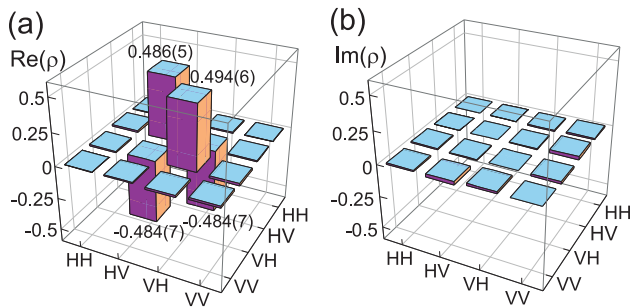}} \caption{(a) Real and (b)
imaginary part of the polarization state density matrix $\rho$ of the photon pairs in the
horizontal-vertical (H-V) polarization basis.
} \label{fig:density}
\end{figure}

To explore the polarization entanglement of the photon pairs, they are split by a 50/50
non-polarizing beam splitter and analyzed by polarization state analyzers in front of the
fiber couplers. We perform a full polarization state tomography measurement following
\cite{White}. After subtracting the accidental coincidences, we reconstruct the pair
photon density matrix $\rho$, as shown in Fig.~\ref{fig:density}. As a measure of the
entanglement quality we calculate the concurrence $C = 0.948\pm0.015$. The overlap
fidelity $F=\langle\Psi^-|\rho|\Psi^-\rangle$ with the maximally entangled singlet state
$|\Psi^-\rangle=\frac{1}{\sqrt{2}}(|$H$\rangle|$V$\rangle-|$V$\rangle|$H$\rangle)$,
amounts to $0.976\pm0.011$. Another practical figure-of-merit are the visibilities of the
polarization anti-correlations in the horizontal-vertical and the $\pm 45 ^{\circ}$
polarization bases, which we find to be $V_{HV} = 99.1\pm0.9\%$ and $V_{\pm} =
97.5\pm0.9\%$.

In conclusion, we have set-up and characterized a narrow-bandwidth, tunable, high
spectral density photon pair source, which will allow for efficient coupling to single
trapped Ca$^+$ ions. The temporal correlation of the photon pairs permits studying the
interaction of single atoms with heralded single photons. Exploiting the high-purity
entanglement of the source, the coupling of photonic and atomic qubit pairs will be
feasible. Prospective experimental applications are entanglement distribution
\cite{Kraus} and quantum repeaters \cite{Briegel}.

\vspace{2mm} This work has been partially supported by the European Commission (SCALA,
contract 015714; EMALI, MRTN-CT-2006-035369), by the Spanish MEC (QOIT, CSD2006-00019;
QLIQS, FIS2005-08257; QNLP, FIS2007-66944; FLUCMEM, FIS2005-03394; ILUMA, FIS2008-01051),
and by the Generalitat de Catalunya (2005SGR00189). A.H. acknowledges support by the
'Juan de la Cierva' and N.P. by the FPU fellowship programme of the Spanish MEC.


\begin{thebibliography}{99}
\bibitem{Carlota} The crystal was fabricated by C. Canalias
from the Laser physics department of the KTH in Stockholm.
\bibitem{footnote_rate}
We could not measure this rate directly, due to saturation of the detector in the
unfiltered arm.
\end{thebibliography}
\end{document}